\documentclass[twocolumn,amsmath,amssymb,pra,nolongbibliography]{revtex4-2}

\usepackage{bm}
\usepackage{amsmath}
\usepackage{dsfont}
\usepackage{physics}
\usepackage{graphicx}

\begin{document}

\title{Demonstration of teleportation across a quantum network code}
\author{Hjalmar Rall}
\email{hjalmar.rall1@gmail.com}
\author{Mark Tame}
\affiliation{
 Department of Physics, Stellenbosch University, Matieland 7602, South Africa
}

\date{\today}

\begin{abstract}
  In quantum networks an important goal is to reduce resource requirements for the transport and communication of quantum information. Quantum network coding presents a way of doing this by distributing entangled states over a network that would ordinarily exhibit contention. In this work, we study measurement-based quantum network coding (MQNC), which is a protocol particularly suitable for noisy intermediate-scale quantum devices. In particular, we develop techniques to adapt MQNC to state-of-the-art superconducting processors and subsequently demonstrate successful teleportation of quantum information, giving new insight into MQNC in this context after a previous study was not able to produce a useful degree of entanglement. The teleportation in our demonstration is shown to occur with fidelity higher than could be achieved via classical means, made possible by considering qubits from a  polar cap of the Bloch Sphere. We also present a generalization of MQNC with a simple mapping onto the heavy-hex processor layout and a direct mapping onto a proposed logical error-corrected layout.  Our work provides some useful techniques for testing and successfully carrying out quantum network coding.
\end{abstract}

\maketitle{}
\section{Introduction}
Quantum communication enables interactions between physically distant quantum systems and opens the path to applications such as distributed quantum computing, quantum key distribution (QKD), and communication within quantum processors \cite{VanMeter2016, Gisin2007, Wehner2018}. Significant progress towards practical quantum communication has been made in recent years: high fidelity quantum communication between a single source and destination has been realized in a number of experiments, including free-space quantum communication \cite{Jin2010}, a large scale QKD network with satellite link \cite{Chen2021}, a 3-node quantum network utilising solid-state qubits  \cite{Pompili2021}, and research into optical communication between superconducting quantum computers is ongoing \cite{Magnard2020, Mirhosseini2020}. These systems are too small at present to be of practical use, but larger and more complex networks are becoming feasible, necessitating the study of quantum networks in a practical context. Of equally great importance are internal communication networks inside quantum processors, as the superconducting processors of Google, IBM, and Rigetti are rapidly growing to sizes where the standard entanglement and qubit swapping approaches become impractical \cite{Das2022}. In addition, methods for external communication networks that enable the linking up of small quantum processors to make an effective larger processor have started to gain attention recently~\cite{Tham22,Buessen22}, which is relevant in light of computational techniques for distributed processors such as entanglement forging~\cite{Eddins22}.  \\

Quantum networks have been studied at length in the literature \cite{Perseguers2013, Chiribella2009, Kimble2008}, but many practical issues remain, especially in the current noisy intermediate-scale quantum (NISQ) era \cite{Preskill2018}, where entanglement is imperfect and many rounds of purification may be required to achieve a sufficient degree of entanglement. Thus the pre-shared entanglement required for a teleportation-based network is difficult to establish and the bandwidth of the network may be severely limited. Quantum networking is further restricted by the need for quantum routing which takes up valuable resources in terms of the number of qubits required and also introduces additional noise into the system. It is therefore necessary to find efficient schemes for quantum networking with limited qubit number and bandwidth. A solution to this is provided by quantum network coding (QNC) \cite{Hayashi2007a, Kobayashi2009}. In classical networks with limited bandwidth, network coding \cite{Ahlswede2000, Ho2008, Li2003} solves the problem of contention by encoding messages which must pass through bottlenecks and using uncontended channels to send decoding instructions. In certain communication scenarios classical network coding can utilise all available bandwidth for useful communication despite the presence of bottlenecks. QNC mimics the classical case in that it makes use of local operations to achieve simultaneous transmission of messages through a bottleneck in a quantum network. In contrast to the classical case, this is achieved by redistributing the available channels (entanglement) so as to eliminate the bottleneck entirely.\\

QNC has experienced considerable interest since its introduction in 2007 \cite{Hayashi2007}, and has been studied both as a theoretical tool and as a practical protocol in quantum networks and processors \cite{Kobayashi2009, Satoh2012}. It has also recently been demonstrated experimentally in an optical setup \cite{Lu2019}. Measurement-based quantum network coding (MQNC) \cite{Matsuo2018} is a very recent development which is well-suited to the NISQ regime by virtue of requiring shallower circuits than existing QNC protocols. As a result of shorter circuit depth, the effect of qubit loss, gate errors, and qubit decoherence is reduced. MQNC has previously been studied on an IBM Q superconducting processor by Pathumsoot et al. \cite{Pathumsoot2020}, but the study was severely limited by the high degree of noise in the processor and did not realize quantum communication over the network code. In the time since, IBM has made new processors available, with greatly reduced noise, and a standardised layout which it is said will remain fixed for the forseeable future. It is of interest then, to see how MQNC performs on this new hardware, and what further insight can be gained into practical implementaiton of MQNC beyond general predictions made in the previous work. We overcome the challenges of translating MQNC to the new processor layout, and show that - even with the extra overhead incurred - genuine quantum information transfer using teleportation over an MQNC network is possible on these processors, provided that the input states are restricted to a polar cap of the Bloch sphere, as in a recent theory proposal by Roy et al. \cite{Roy2022}.\\

With a view to the future, we also present a generalisation of butterfly MQNC to a non-blocking network switch with an arbitrary number of nodes. Interestingly, the switch may be created directly on square grid topologies which are already in use or planned for use in superconducting quantum processors - examples include the Google Sycamore processor \cite{Arute2019} and the planned error-corrected logical topology of the IBM processors \cite{Chamberland2020}. While this is important for transferring quantum information within processors, it also has implications for networking within a quantum internet, where switching within quantum routers \cite{Huberman2020} is essential if entanglement between arbitrary pairs of end nodes is to be established using a shared physical link layer instead of private direct connections between nodes.\\

The paper is structured as follows: In section II we give a brief overview of measurement-based quantum computing, QNC, and MQNC, and introduce the generalised switch. In section III we present the particulars of the protocol and the method used to adapt the previous work to a newer processor. Section IV forms the main body, where the results of teleportation using MQNC are presented. Section V presents a general mapping of MQNC onto IBM processors. We end with a discussion and concluding remarks in Section VI.

\section{Background}
\subsection{Measurement-Based Quantum Computing}
A graph state is an entangled state $\ket{G}$ with qubits and entanglement between qubits corresponding to the vertices and edges of an undirected graph $G = (V, E)$. An $N$-qubit graph state with edge set $E$ is defined according to
\begin{equation}
\ket{G} = \prod_{\{i,j\}\in E}CZ_{i,j}\ket{+}^{\otimes N} ,
\label{eq:graphstate}
\end{equation}
where $CZ_{ij}$ is the controlled phase operation on qubits $i$ and $j$, and $\ket{+}$ is the Pauli-$X$ eigenstate with eigenvalue +1. A number of quantum operations transform between graph states, and can thus be viewed as operations on the underlying graph. The following are the most common transformation rules that will be used in this work \cite{Hein2004}:\\
\begin{itemize}
        \item \textbf{T1:} A Z-basis measurement on a qubit $a$ removes the corresponding vertex and incident edges from the graph.
        \item \textbf{T2:} A Y-basis measurement on a qubit $a$ removes the corresponding vertex and incident edges, and complements the subgraph induced by the neighbourhood $N_{a}$. i.e. $G(V,E) \rightarrow G(V\slash \{a\}, E \Delta K)$ with $K$ the edge set for the complete subgraph induced by $N_{a}\cup \{a\}$ and $\Delta$ the symmetric difference. In other words, neighbours of $a$ are connected unless a connection already exists, in which case it is broken.
        \item \textbf{T3:} X-basis measurements on two adjacent qubits $a$ and $b$ removes them and complements the bipartite subgraph induced by $N_{a}$ and $N_{b}$. i.e. $G(V,E) \rightarrow G(V\slash \{a,b\}, E \Delta K)$ with $K$ the complete bipartite subgraph induced by $N_{a}\cup \{a\}$ and $N_{b}\cup \{b\}$. In other words, all the neighbours of $a$ are connected to all the neighbours of $b$ unless a connection already exists, in which case it is broken.
        \item \textbf{T4:} Local complementation on a qubit $a$ given by $\sqrt{X_{a}}\prod_{b\in N_{a}} \sqrt{Z_{b}}\ket{G}$ complements the subgraph induced by $N_{a}$ and leaves $a$ and its incident edges unchanged. In other words, neighbours of $a$ are connected unless a connection already exists, in which case it is broken. This is a non-destructive version of the Y-measurement transformation rule T2 where the qubit $a$ is not removed from the graph.
\end{itemize}

\begin{figure}
	\centering
  \includegraphics[width=0.45\textwidth]{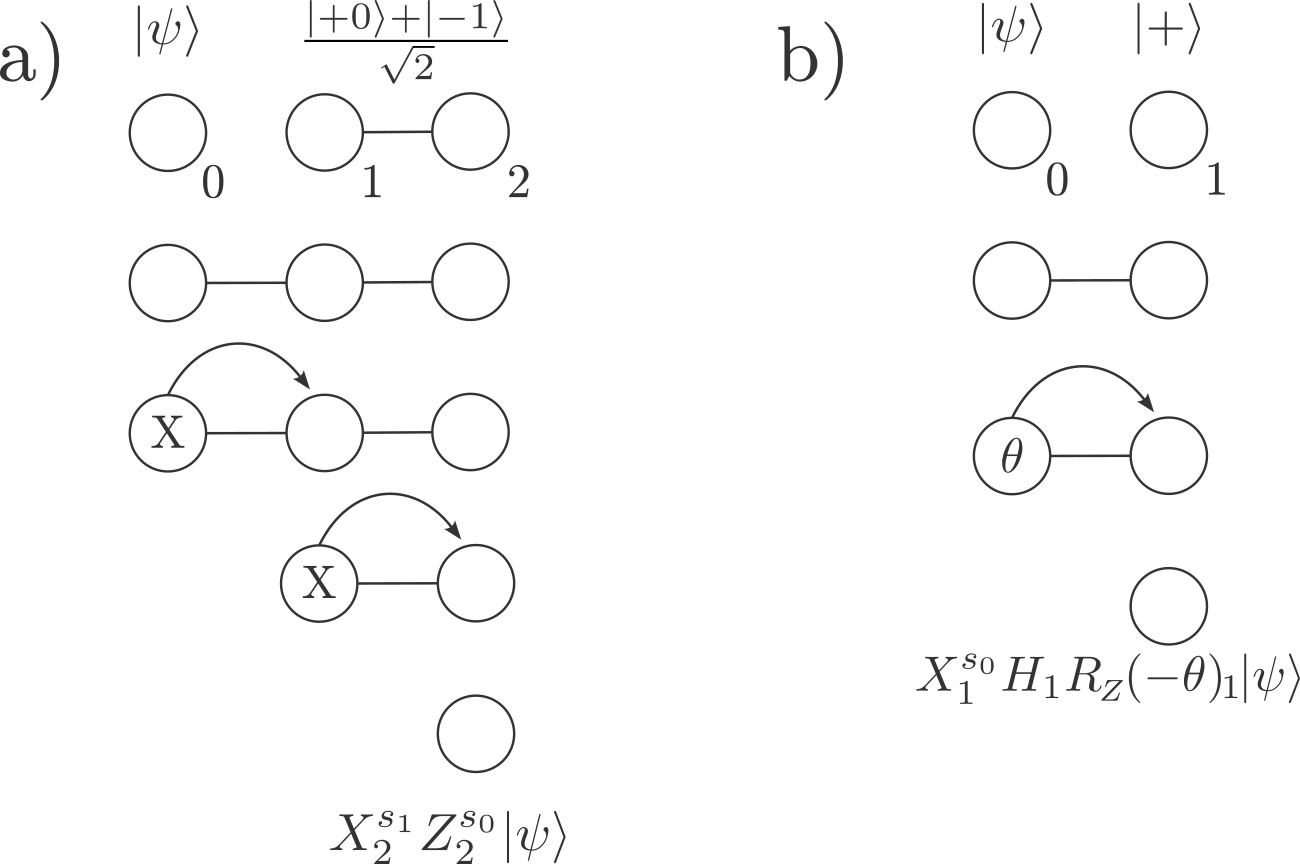}
  \caption{\label{fig:teleportation} (a) A state \(\ket{\psi}\) may be teleported across a linear section of a graph state. (b) Teleportation can be used to perform the operation \(HR_{Z}(-\theta)\). \(X\)'s within qubits indicate \(X\)-measurements, \(\theta\) indicates a measurement in the basis \(\{\frac{1}{\sqrt{2}}(\ket{0}+e^{i\theta}\ket{1}),\frac{1}{\sqrt{2}}(\ket{0}-e^{i\theta}\ket{1})\}\).}
\end{figure}
It should be noted that Pauli byproducts are introduced for certain measurement outcomes for the above operations. These must be tracked by a classical computer and either adaptively corrected or commuted through to the end of the circuit before being corrected (possibly via post-selection). For an in-depth discussion of graph states we refer the reader to Ref. \cite{Hein2004}. Graph states form a resource for measurement-based quantum computation (MBQC) \cite{Raussendorf2001, Briegel2009}, many elements of which are used in MQNC. An arbitrary single-qubit state $\ket{\psi}$ may be attached to a graph state using a controlled phase gate. This state may subsequently be transported within the graph by means of quantum teleportation along linear sections. An example of this is shown in Fig.~\ref{fig:teleportation}(a): First a 2-qubit (linear) graph state \(\frac{1}{\sqrt{2}}(\ket{+0} + \ket{-1})_{12}\) is created, then the state \(\ket{\psi}_0\) to be teleported is entangled with this state via a controlled phase gate, and finally the measurements \(M_{x,0}\) and \(M_{x,1}\) in the basis \(\{\frac{1}{\sqrt{2}}(\ket{0}+\ket{1}),\frac{1}{\sqrt{2}}(\ket{0}-\ket{1})\}\) with outcomes \(s_0\) and \(s_1\), respectively are performed. This yields the state \(X_2^{s_1}Z_2^{s_0}\ket{\psi}_2\) which may be transformed back to \(\ket{\psi}\) if the measurement outcomes are known. 

Arbitrary unitary operations may be performed on a single-qubit state attached to a graph state by way of appropriate projective measurements, which serve to both teleport and transform the single-qubit state. An example is shown in Fig.~\ref{fig:teleportation}(b): First the state \(\ket{\psi}_0\) is entangled with the graph state (here the single qubit graph \(\ket{+}_1\)) via a controlled phase gate, then a measurement in the basis \(\{\frac{1}{\sqrt{2}}(\ket{0}+e^{i\theta}\ket{1}),\frac{1}{\sqrt{2}}(\ket{0}-e^{i\theta}\ket{1})\}\) (with \(\theta\) an arbitrary angle) is performed on qubit 0. This yields the state \(X_1^{s_0}H_1R_Z(-\theta)_{1}\ket{\psi}_{1}\). Such operations may be composed to obtain arbitrary unitary operations. Given a sufficiently large 2D grid (cluster) graph state, measurement-based computation is universal. Since all two-qubit operations are performed during the creation of the resource state, and they are all commutative, they may be done simultaneously if the hardware allows. Since all byproducts are Pauli operators, they can be commuted through to the end of the circuit either directly or adaptively and subsequently simplified, leading to further decreases in circuit depth.
\begin{figure}
  \centering
  \includegraphics[width=0.4\textwidth]{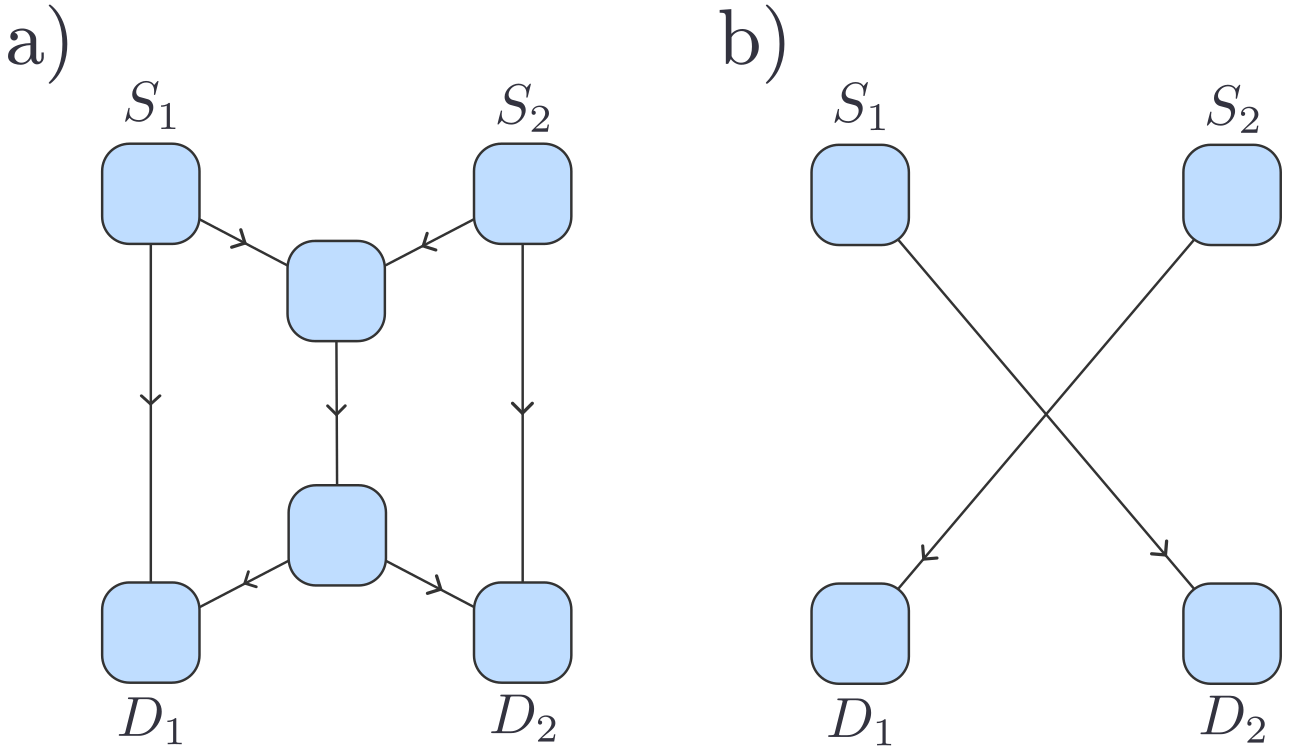}
  \caption{\label{fig:qnc} The butterfly network (a), and the new network after performing QNC (b). Blue shading represents network nodes. Lines represent quantum communication channels. Classical communication is assumed to be free.}
\end{figure}

\begin{figure}
  \centering
  \includegraphics[width=0.4\textwidth]{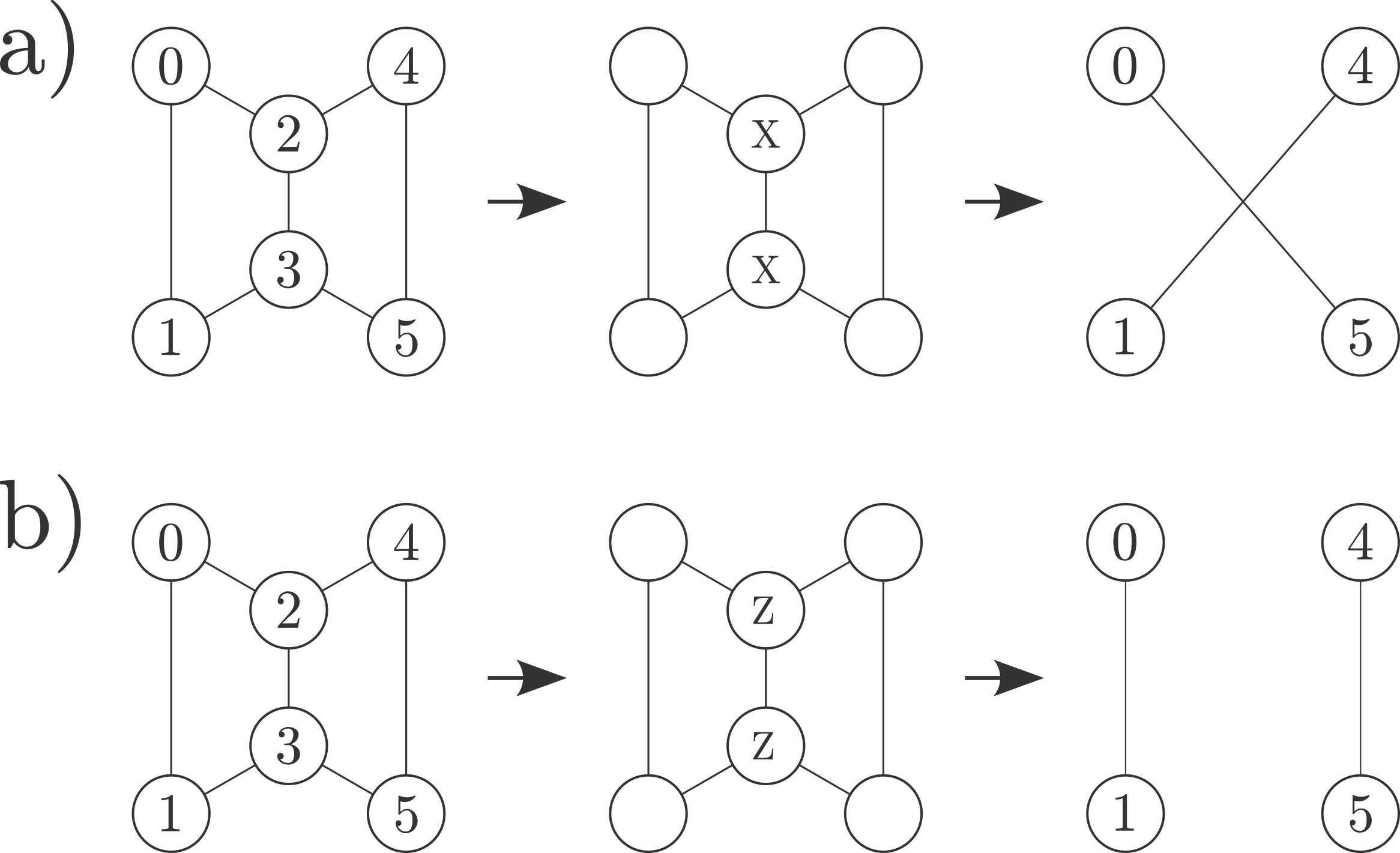}
  \caption{\label{fig:mqnc} The procedure for MQNC starting with a 6-qubit graph state. (a) and (b) show the two different configurations (cross pairs and straight pairs) of MQNC. Circles and lines (which are vertices and edges of graphs, respectively) represent qubits and entanglement between qubits.}
\end{figure}

\begin{figure*}[tb]
  \centering
  \includegraphics[width=0.85\textwidth]{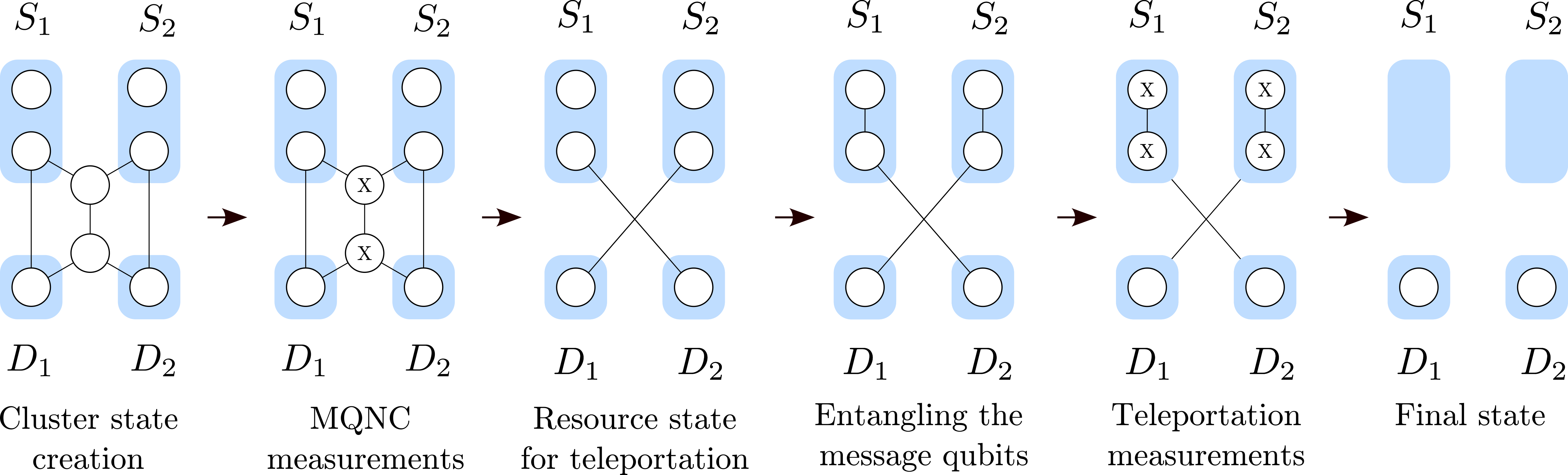}
  \caption{\label{fig:mqncteleport} The procedure for performing MQNC to generate cross pairs and subsequently teleporting states across these pairs. The straight pair (Z-measurement) case outlined in Fig.~\ref{fig:mqnc}(b) is similar. White circles represent qubits and blue shaded regions indicate a single source or destination node in the network. }
\end{figure*}

\subsection{Quantum Network Coding}
Quantum network coding is best illustrated through the example of the butterfly network. Given the network shown in Fig.~\ref{fig:qnc}(a) with each channel having capacity 1, the goal is to simultaneously send qubits from $S_1$ to $D_2$ and from $S_2$ to $D_1$. In Ref. \cite{Hayashi2007} it is shown that it is not possible to do this perfectly if only quantum communication is allowed. It was later shown  that perfect quantum network coding is possible if free classical communication is allowed \cite{Kobayashi2009}, as shown in Fig.~\ref{fig:qnc}(b). Protocols for perfect QNC have been developed for the case where transmitters share entanglement \cite{Hayashi2007a} and for the case of the butterfly network across quantum repeaters \cite{Satoh2012}. The former protocol has been demonstrated experimentally in an optical setup \cite{Lu2019} with fidelity sufficient to enable teleportation of quantum information with fidelity exceeding the classically achievable bound. These protocols however require complex circuits and additional steps for resource state creation. On the other hand, MQNC \cite{Matsuo2018} presents a measurement-based alternative to the repeater network protocol with a reduction in circuit depth of 50\% and a corresponding increase in the allowable gate error to achieve a specified fidelity. Furthermore, this protocol contains as an intermediate step a graph state which also has applications in on-processor teleportation.\\

The protocol proceeds as follows:
Starting with seven bell pairs, a six-qubit graph state is generated as shown in Fig.~\ref{fig:mqnc}(a) on the left hand side. This state may also be generated directly via controlled phase gates according to Eq. \eqref{eq:graphstate}. By measuring the two central qubits in the $X$-basis (Fig.~\ref{fig:mqnc}(a) middle) the entanglement in the graph state can be redistributed so as to give two cross pairs (up to Pauli byproducts) using transformation rule T3, or alternatively, by measuring the central qubits in the $Z$-basis, the state shown on the right-hand side of Fig.~\ref{fig:mqnc}(b) is created (up to Pauli byproducts) using transformation rule T1. The 2-qubit graph states generated by the MQNC measurements are maximally entangled and can be used for teleportation, where qubits are first entangled with the state generated by MQNC and subsequently teleported to the destination nodes, as shown in Fig. \ref{fig:mqncteleport}. Here, the states to be teleported may also be entangled before the MQNC protocol begins, as the entangling commutes with the measurements of the central qubits. Effectively, the three vertical channels with a bottleneck along one of the possible desired routes have been used to create two possible configurations through which the desired quantum communication routes may be achieved directly.\\

\subsection{Generalized MQNC}
\begin{figure*}[!tb]
\includegraphics[width=0.9\textwidth]{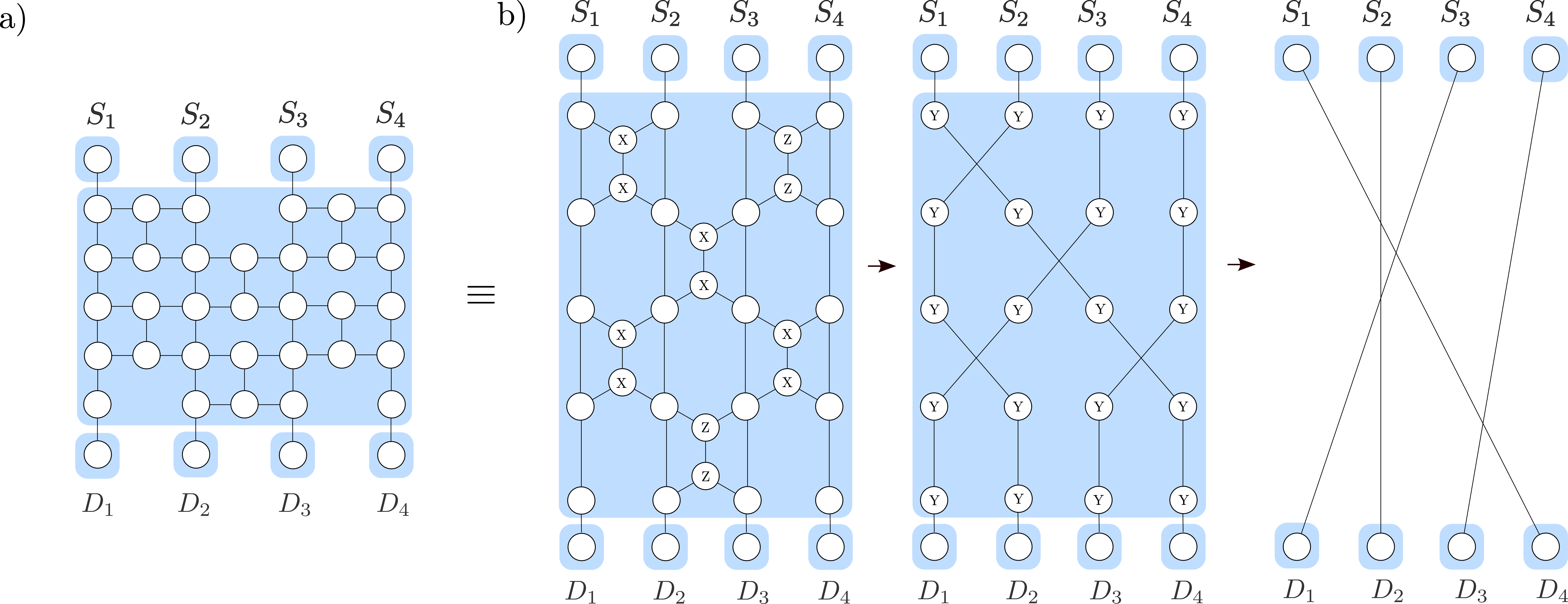}
\caption{\label{fig:switch} A measurement-based quantum switch utilising MQNC. A Spanke-Bene\v{s} network is chosen because it is optimal for planar networks. The square grid structure is showcased in the $4\times 4$ switch shown in (a). In (b) the same network is expanded to show how it is made up of butterfly networks. The measurements in (b) give an example of how MQNC can be used to achieve a particular permutation of connections between 4 sources and 4 destinations. After performing the MQNC measurements, Y-measurements are used to remove residual qubits according to transformation rule T2.}
\end{figure*}

MQNC has thus far been considered in the same context as traditional classical network coding, where the primary goal is to increase network throughput. Quantum networks however are limited to the multiple unicast scenario ($k$-pair problem \cite{Leung2010, Li2017}) due to no-cloning \cite{Wootters1982}, and therefore the primary benefit of QNC is that it solves contention and therefore succeeds in limited capacity networks where quantum routing fails. The problem of communicating within a quantum network ultimately reduces to simultaneously connecting $k$ pairs of source and destination qubits. This is more closely related to switching than to routing, and indeed butterfly QNC acts like a $2\times 2$ non-blocking switch.\\

Motivated by this interpretation, before proceeding with the details of our demonstration, we present a generalization of MQNC in Fig. \ref{fig:switch}. Unlike the previous generalization in Ref. \cite{Li2017}, our scheme requires no more than four connections per qubit regardless of the number of source and destination nodes. We have not seen such a construction in the literature \cite{Hayashi2007, Kobayashi2009, Matsuo2018}. A similar quantum switching network was studied in Ref. \cite{Ratan2007}, but the $2\times 2$ switches in that network were based on CNOT and Toffoli gates and thus differ substantially from the measurement-based approach presented here. Our switching network consists of multiple copies of the MQNC code joined together as shown in Fig.~\ref{fig:switch}(a) in a Spanke-Bene\v{s} network \cite{Spanke1987}. This network is chosen because it is an optimal non-blocking planar network, the planar property being essential for 2-dimensional quantum processor topologies which cannot directly implement crossed connections as in a standard non-planar Bene\v{s} network. The network coding measurements create linear graph states between source and destination as shown in Fig.~\ref{fig:switch}(b), and the intermediate qubits in these states are subsequently removed via Pauli-Y measurements using transformation rule T2. If the switch is centrally located, and each source or destination qubit is at a separate location, only $2k$ non-local connections to the switch are required.\\

Our generalised MQNC scheme uses a total of $\frac{k(k-1)}{2}$ switches, which was shown in Ref. \cite{Spanke1987} to be the smallest number for a non-blocking planar network. Since there is some overlap between switches, the total number of qubits required is only $4k(k-1) + k$. The number of $2\times 2$ MQNC switches can be reduced while keeping the non-blocking property by making use of an arbitrary size Bene\v{s} network \cite{Chang1997}, which uses only $O(k \log_2 k)$ $2\times 2$ switches, but this might be difficult to implement on 2-dimensional quantum processor topologies since Bene\v{s} networks \cite{Benes1964} have crossed connections (i.e. they are non-planar). The Spanke-Bene\v{s} network on the other hand has a structure which is well suited to these topologies, and in fact it may be directly created given a square grid topology. We show later how it may also be created easily on the so-called `heavy-hex' topology currently in use on IBM quantum processors.

\section{MQNC on IBM processors}
Previously, Pathumsoot et al.  \cite{Pathumsoot2020},  studied MQNC on the superconducting processor IBM Q 20 Tokyo by generating the 6-qubit resource graph state shown in Fig.~\ref{fig:mqnc} (left hand side) directly using the two-qubit gates available on the processor. They showed that entanglement existed in the resource state, and reported on the fidelity and concurrence of the final 2-qubit states, achieving fidelities of $0.57\pm0.01$ and $0.58\pm0.01$ for the cross pairs (fidelities for straight pairs were not reported). Lastly, they tested for violation of the CHSH inequality in the final states, but found none due to deterioration of the state from processor noise. It was determined that CHSH violation should be possible with single qubit gate errors approximately half those of IBM Q Tokyo.

The IBM superconducting processors are devices undergoing rapid development, and the past two years have seen dramatic decreases in noise and errors. One such processor is ibm\_cairo, a 27 qubit device of the `falcon' series. As shown in Table \ref{table:cairo} in the appendix, it has significantly reduced error compared to IBM Q 20 Tokyo, the error rates for which are reported in Table \ref{table:tokyo} in the appendix. As with all current IBM processors, however, this reduction in error rates comes at the cost of a processor topology wherein the harware for two-qubit operations is more sparsely distributed among the qubits. While this processor topology has definite advantages and is planned to be IBM Quantum's standard for the forseeable future, it precludes the possibility of creating the 6-qubit resource state using direct entangling operations, as was done with a specialized processor topology in \cite{Pathumsoot2020}. In light of this, we consider it prudent to adapt the demonstration of Pathumsoot et al. to these newer processors, to show that the connectivity limitations of the improved processors can be overcome, and subsequently to show stronger evidence that MQNC will become practical on near-future NISQ devices.

We performed MQNC as outlined in Fig.~\ref{fig:mqncteleport} on the IBM Q falcon superconducting processor ibm\_cairo.  The processors were accessed remotely through the IBMQ API and the QISKIT Python library. In place of standard circuit transpiling techniques utilising excessively noisy SWAP operations we employ graph state rewiring techniques in order to overcome the processor topology limitations; a discussion of this follows in the next subsection. We subsequently demonstrate that the resource required for performing MQNC can be generated with fidelity exceeding that obtained in the prior work. Despite this, it is still not possible to perform full teleportation across the resource using MQNC. However, we show that by considering the teleportation of a region of states from the Bloch sphere the utilisation of quantum correlations during the teleportation can be confirmed, thus demonstrating for the first time the use of quantum effects in MQNC.

\subsection{Processor topology}
\begin{figure}
  \includegraphics[width=0.48\textwidth]{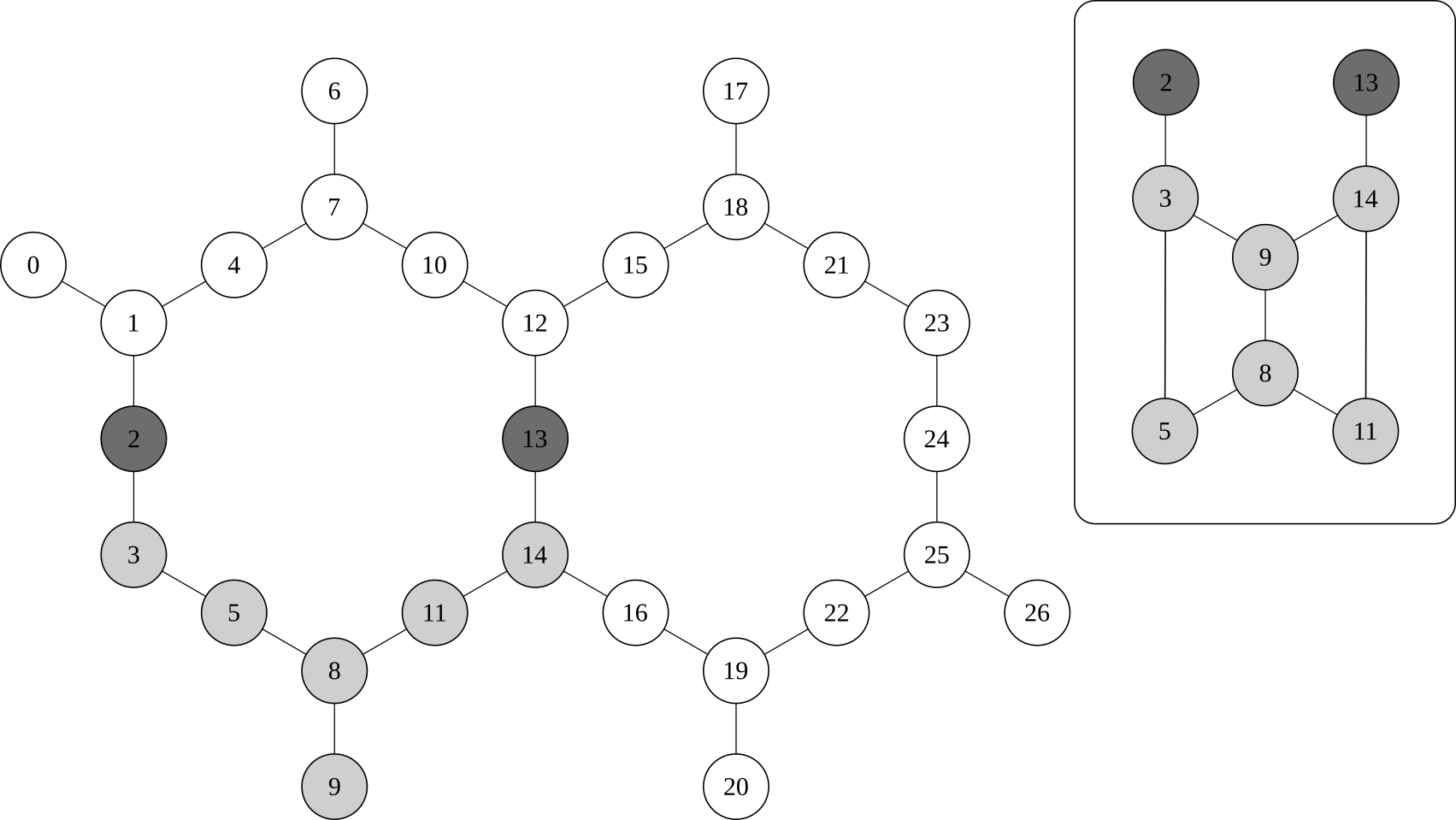}
  \caption{Layout of ibm\_cairo. Edges indicate neighbouring qubits on which CNOT gates may be performed. Light grey qubits are used to create the 6-qubit graph state for MQNC. Dark grey qubits are teleported across the network code graph state by entangling them with the graph state and subsequently performing appropriate measurements. The inset shows the new layout after rewiring has been performed.}
  \label{fig:layout}
\end{figure}
All current IBM Q processors make use of variations on the layout shown in Fig.~\ref{fig:layout}. This layout was chosen in anticipation of future generations of the processors which will have sufficiently low gate error to implement error correction codes (ECCs) native to this layout \cite{Chamberland2020}. These are stabilizer codes with logical qubits arranged in a square lattice, where two-qubit operations may be performed between nearest neighbours. Consequently they would also allow for the creation of error-corrected or `encoded' graph states based on the square grid. It is interesting to note that - since graph states are stabilizer states - such an encoded graph state may be viewed as a single stabilizer state of the physical qubits which encodes multiple logical qubits in its stabilized subspaces. This is related to the work in Refs. \cite{Schlingemann2001, Schlingemann2002, Grassl2002}, where it was shown that there is a one-to-one correspondence between stabilizer codes and graph codes.

In current processors, where the gate error is still too high for implementation of the ECCs in Ref. \cite{Chamberland2020}, physical qubits must be used, and we are limited by the reduced connectivity of the heavy-hex lattice (compared to the square lattice) which is not well suited to the creation of arbitrary graph states. Nonetheless, direct graph state generation on the heavy-hex lattice is possible. Non-adjacent qubits may be entangled by use of SWAP operations, and the QISKIT transpiler provides functionality to determine the necessary operations automatically and perform optimization. While the QISKIT transpiler works in general, it fails to produce a practical circuit for the generation of a 6-qubit graph state, as the large number of 2-qubit gates introduces an excessive amount of noise. During the initial stages of our study, we attempted to generate the 6-qubit graph state using SWAP-based transpiling. However, we found that this resulted in states close to completely mixed states for the resulting pairs due to the large number of 2-qubit gates required that introduce an excessive amount of noise. More advanced transpilers are available, see for instance those used in Refs. \cite{Childs2019} and \cite{Li2019}, but these are also not tailored to graph-state generation.

For example, in order to entangle an arbitrary pair of qubits it is first necessary to apply swap operations until the states occupy adjacent physical qubits. Doing so incurs a significant amount of noise as each swap operation consists of three controlled-not gates with an average error of the order of $4\times 10^{-2}$. Furthermore, the calculation of an optimal sequence of swap gates incurs a non-trivial classical processing cost. Attempting to transpile the MQNC circuit to the processor using the built-in tools of the QISKIT package results in a circuit incorporating a minimum of twelve noisy 2-qubit gates, and results in final states close to the maximally mixed state. Effectively, the new processors are no better than the old one at performing MQNC when this transpiling scheme is used. We present here an alternative transpiling scheme based on local complementation (transformation rule T4) which requires only seven 2-qubit gates, and yields a circuit of reasonable depth. This means that the transpiling step, while still noisy, is no longer so noisy as to completely negate the improvements in hardware in the new processors.\\

Let $a$ be a qubit in a graph state $|G\rangle$ with graph $(G,E)$. Let $N_a$ be the neighbourhood of $a$ in $|G\rangle$. The local operation
$LC_{a} = \sqrt{X_a}\prod_{b \in N_a}\sqrt{Z_b}|G\rangle,$ known as local complementation produces a new graph state with graph $G^\prime(V, E^{\prime})$ with $ E^{\prime}= E\Delta E(N_a, N_a)$ following transformation rule T4. As an example of using local complementation to redistribute entanglement, consider three qubits $a$, $b$, and $c$ in a graph, of which only $a$ and $b$, and $b$ and $c$ may be entangled directly via an edge. Entanglement between $a$ and $c$ can be created by performing local complementation on $b$. This leaves all three qubits entangled. Any one edge in this new triangle graph may be removed by performing local complementation on the opposite qubit. Any two edges may be removed by performing a $Z$-measurement on their common qubit which has the effect of deleting all incident edges on that qubit following rule T1. In this way an operation requiring six 2-qubit operations using swap gates is achieved using only two with local complementation.

\begin{figure}
  \includegraphics[width=0.4\textwidth]{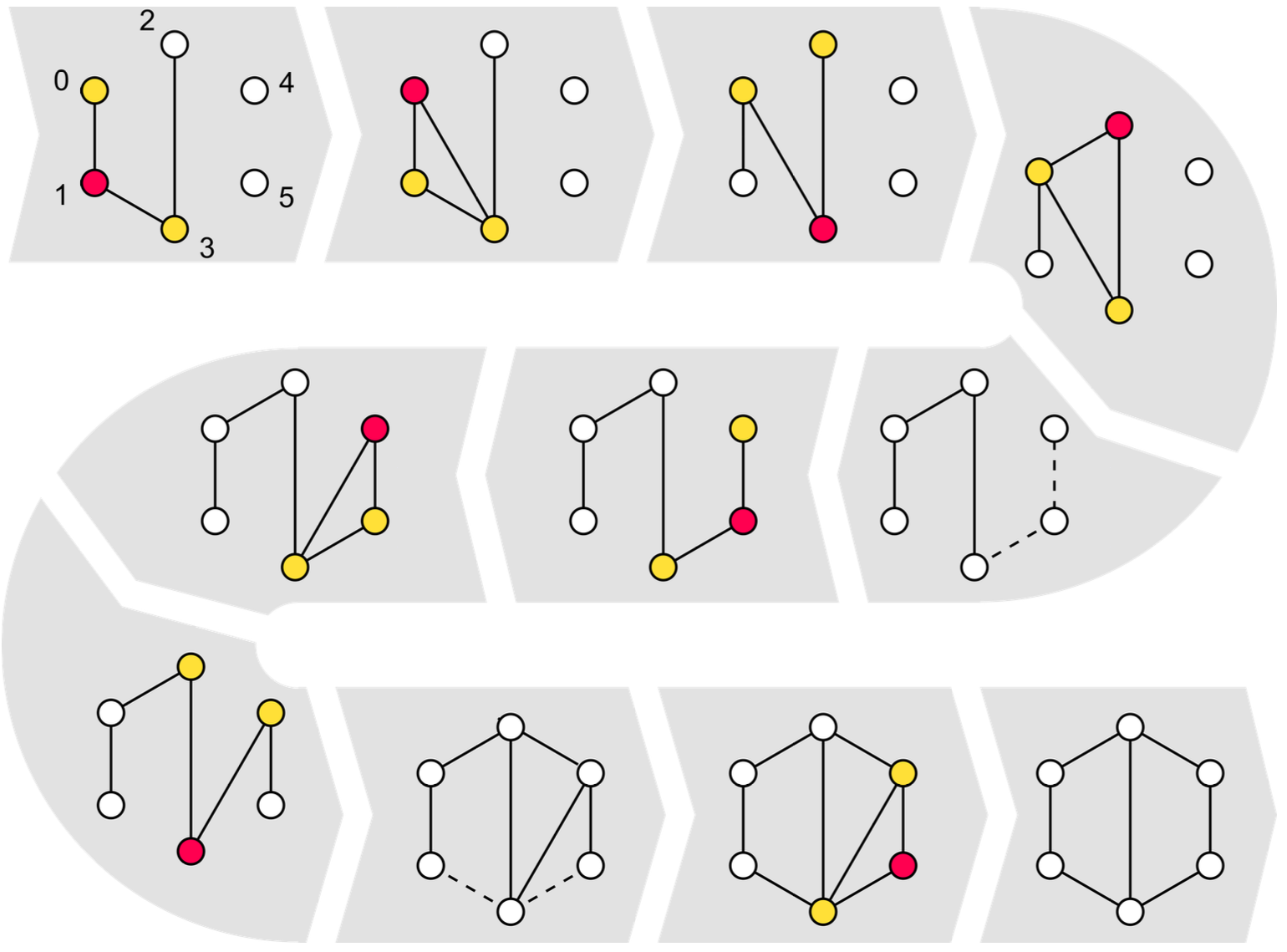}
  \caption{The procedure for ``rewiring'' the processor layout to obtain the 6-qubit resource state for the network code. We start by creating the three edges in the first graph. Local complementation is performed on pink qubits. Yellow qubits are in the neighbourhood of the latter. Dashed lines indicate further entangling operations.}
  \label{fig:rewire}
\end{figure}

In order to generate the 6-qubit graph state for MQNC, we implement the following sequence of operations (from right to left) which are shown in Fig.~\ref{fig:rewire}:
\begin{align*}
  &LC_{5}CZ_{1,3}CZ_{3,5}LC_{3}LC_{4}\\
  \times &LC_{5}CZ_{3,5}CZ_{4,5}LC_{2}LC_{3}\\
  \times &LC_{0}LC_{1}CZ_{2,3}CZ_{1,3}CZ_{1,0}\\
\end{align*}
where the logical qubits 0, 1, 2, 3, 4, and 5 are mapped onto the qubits 5, 3, 8, 9, 11, and 14 on the processor, as shown in Fig.~\ref{fig:layout}, as these represent the best least error-prone set of qubits with the appropriate shape and having no qubit which is particularly error-prone. Qubits 2 and 13 are then the states to be teleported across the resource.

\section{Implementation on new processors}
We implemented MQNC on ibm\_cairo using the rewiring scheme introduced above. A single run constitutes calibration for readout error mitigation, state tomography of the 6-qubit resource state, state tomography of one of the two qubit graph states generated using MQNC, and process tomography of teleportation across the same two qubit graph state. This is repeated for each of the four pairs in turn, and the whole procedure is then repeated 30 times to account for variation in processor noise. The runs were spread out between 9PM 26 Apr 2022 and 5AM 27 Apr 2022 (UTC) due to use of the fair-share queuing system and the daily calibration data are given in the appendix. The variation in processor noise is found to be sufficiently small over a time-scale of minutes, and that the delay between readout error mitigation calibration and demonstration is not significant, so that results are not skewed from one pair to the next. 

The measurements in the protocol result in probabilistic byproduct operations on the desired final state. At the time of use, the IBM processors did not support the feed-forward functionality necessary to undo these operations (this is expected in the near future), so for testing purposes results must be post-selected based on measurement outcomes. The four possible outcomes of two measurements at the central qubits of the network code correspond to four possible byproducts (including the identity) on each of the final 2-qubit states, which occur with equal probability so that 1/4 of the results are kept. The measurements required for the teleportation also result in byproduct operations, but these are either one of the operations in $\{I, X, Z, XZ\}$ with equal probability so that combined with the byproducts from the network coding we have the identity byproduct on the final teleported qubit with probability 1/4, and 1/4 of the results remain after post-processing.\\

\begin{figure*}
\centering
\includegraphics[width=0.80\textwidth]{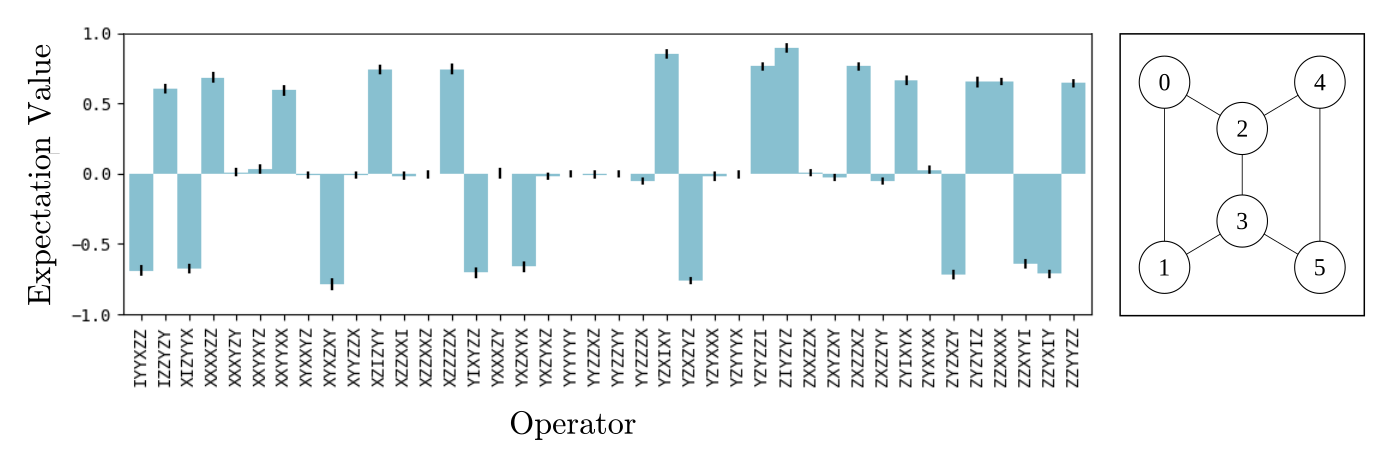}
  \caption{\label{fig:clusterfids} Expectation values of the stabilizers of the 6-qubit graph state resource. The convention used is that the tensor product operators read from left to right with the leftmost acting on qubit 0 and the rightmost on qubit 5.}
\end{figure*}

\subsection{Genuine Multipartite Entanglement}
The demonstration of genuine multipartite entanglement (GME) on IBM processors has proved challenging so far. In particular, a recent study attempting to create a 20-qubit linear graph state on ibmq\_poughkeepsie \cite{Mooney2019} found that no subset of more than 3 qubits had undeniable GME. To confirm that the 6-qubit graph state for MQNC can be generated successfully we perform the measurements necessary to calculate a fidelity-based GME witness \cite{Toth2005}. First note that graph states are stabilizer states with stabilizers for an $N$-qubit state given by $S_{i} = X_{i} \prod_{j=0}^{N-1}Z_{j}^{(\Gamma_{ij})}$, where $\Gamma$ is the adjacency matrix for the corresponding graph. The graph state is uniquely defined by these stabilizers, and explicitly given by $\rho = |G_{2\times 3}\rangle \langle G_{2\times 3}| = \prod_{i=0}^{5}\frac{\mathds{1} + S_{i}}{2}$ \cite{Lu2019}. Multiplying out this product we get a sum over 64 projectors. Given the expectation value of each the fidelity may be calculated and from that it can be determined if GME exists. As in the article by Pathumsoot et al. \cite{Pathumsoot2020}, we define fidelity according to
\begin{align*}
  F(\rho, \sigma) &=  {\left(Tr\left[\sqrt{\sqrt{\rho}\sigma\sqrt{\rho}}\right]\right)}^{2}\\
                    &= \text{Tr}(\ketbra{G_{2\times 3}}\sigma),
\end{align*}
where the final state and expected state have density matrices $\sigma$ and $\rho$, respectively.
Since the expectation value of the identity is simply 1, the expectation value of any projector of \(\rho\) acting as the identity on one qubit of \(\sigma\) may be determined without reference to that qubit. Hence multiple measurements may be combined into one and only 40 local measurement settings are required to obtain the full set of 64 expectation values. 2000 shots are used per measurement basis. The results are shown in Fig.~\ref{fig:clusterfids}. A fidelity of $0.74\pm 0.02$ was obtained. 

T\'{o}th and G\"{u}hne \cite{Toth2005} give the witness $\langle\hat{\mathcal{W}}\rangle = \alpha\mathds{1} - |G\rangle\langle G|$ for GME near a graph state, where $\alpha$ is defined to be the maximum overlap of the state $|G\rangle$ with any bipartition of qubits. If the witness is negative, then the overlap of the state with the ideal one is greater than the overlap of any bipartition with the ideal state, and hence the state is not biseparable. Therefore $\langle\hat{\mathcal{W}}\rangle$ is a GME witness. The routine provided in QUBIT4MATLAB \cite{Toth2008} was used to calculate $\alpha$ and it was found that $\alpha = 0.5$, leading to a negative $\langle\hat{\mathcal{W}}\rangle$ of $-0.24\pm 0.02$ with a large margin below zero. We conclude that the implementation of the 6-qubit graph state has GME and together with the high fidelity this indicates a largely successful creation of the graph state.\\

\begin{figure}
  \includegraphics[width=0.5\textwidth]{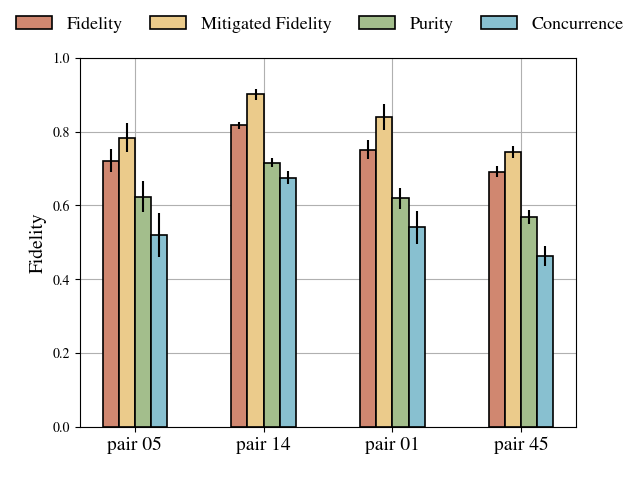}
  \caption{Results of full state tomography on the 2-qubit entangled states generated using the MQNC protocol on ibm\_cairo. Fidelity is shown before and after application of readout error mitigation on the 2 qubits. Purity is defined as \(P=\text{Tr}(\rho^2)\). The protocol qubits 0, 1, 4 and 5 (see Fig.~\ref{fig:clusterfids}) correspond to processor qubits 3, 5, 14 and 11 (see Fig.~\ref{fig:layout}), respectively. Single- and two-qubit error rates are given in Table~\ref{table:cairo} in the Appendix for reference.}
  \label{fig:statetomo}
\end{figure}

\begin{figure*}
  \centering
    \includegraphics[width=\textwidth]{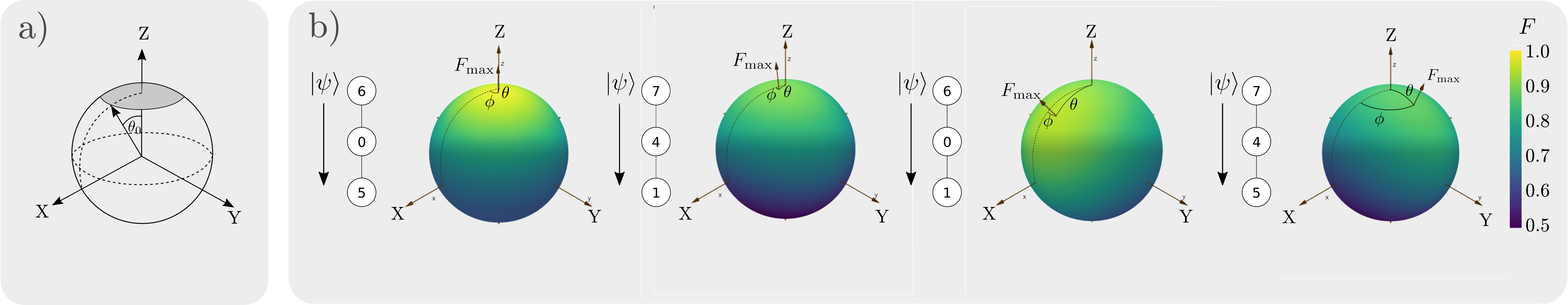}
    \caption{\label{fig:bloch} A polar cap of the Bloch sphere is shown in (a) along with its angular radius $\theta_{0}$. (b) Shows the teleportation fidelity as a function of initial state for each of the four teleportation routes. The highest fidelities fall within a spherical cap of the Bloch sphere (i.e. a polar cap rotated through the angles $\theta$ and $\phi$). The arrow $F_{\text{max}}$ crosses the surface at the point of highest fidelity. The state to be teleported is initially encoded in the qubit labelled 6 (for the case of 05 and 01) or 7 (for the case of 41 and 45), where the protocol numbering from Fig. \ref{fig:mqnc} is used, rather than the processor numbering.}
\end{figure*}

\subsection{2-qubit State Tomography}
In order to measure the quality of the state generated and to compare to the results of the previous study \cite{Pathumsoot2020}, we perform quantum state tomography \cite{James2005} on each of the four pairs that can be generated using MQNC, as shown in Fig.~\ref{fig:mqnc}(a) and (b) on the right hand side. Readout error mitigation is then applied to the results, and they are subsequently post-selected so as to correspond to the byproduct-free graph state. 4000 shots are used for each of the 9 tomography circuits so that approximately 1000 shots remain after post-selection. Density matrices are obtained from the tomography results, and the fidelity compared to the ideal state $\ket{G_{2}} = \frac{1}{\sqrt{2}}(|0+\rangle + |1-\rangle)$, as well as the purity and concurrence are calculated. Concurrence is an entanglement monotone defined according to
\[C(\rho) = \max\{0,\lambda_1- \lambda_2- \lambda_3- \lambda_4\},\]
where the $\lambda_i$ are the square roots of the eigenvalues of \(\tau=\rho(Y\otimes Y)\rho^{*}(Y\otimes Y)\) in descending order.
The results are shown in Fig.~\ref{fig:statetomo}. It is interesting to consider a potential connection between the quality of the pairs and the error rates of the corresponding qubits of the processor. The `protocol’ qubits 0, 1, 4 and 5 (see Fig.~\ref{fig:clusterfids}) correspond to `processor’ qubits 3, 5, 14 and 11 (see Fig.~\ref{fig:layout}), and from best to worse for the fidelity of the protocol qubit pairs in Fig.~\ref{fig:statetomo} we have: (1,4), (1,0), (0,5) and (4,5). From this, we may infer that protocol qubit 5 (processor qubit 11) may be the worst qubit, followed by protocol qubit 0 (processor qubit 3), then protocol qubit 4 (processor qubit 14), and finally protocol qubit 1 (processor qubit 5). In other words, in terms of a contributing factor to the low fidelity of the pairs we have for the processor qubits the hierarchy: $5 < 14 < 3 < 11$. However, by analyzing the error rates in Table~\ref{table:cairo} of the Appendix and noting the single-qubit error rates, we have the error-rate hierarchy $14 < 3 < 11 < 5$, which does not match up with the fidelity of the pairs. 

On the other hand, the two-qubit CX gate error for processor qubit pair (11,14) is the worst and it may explain why this pair (protocol qubit pair (4,5)) is the worst. But, due to the local complementation sequence used to generate the graph state, shown in Fig.~\ref{fig:rewire}, the impact of this 2-qubit gate is not easy to track. However, it could be a main contributing factor. Further detailed work on the propagation of noise would help in assessing how different single-qubit and two-qubit error rates impact the fidelity of the final pairs in the network code using the generation method we have proposed.

Despite the noise incurred by the entire rewiring process, ibm\_cairo shows a significant improvement in both fidelity and concurrence over the results of Pathumsoot et al. \cite{Pathumsoot2020}. Notably, the cross pairs show an improvement in concurrence by a factor of 2 compared to \(0.25 \pm 0.02\) and \(0.36 \pm 0.02\) respectively obtained by Pathumsoot et al.. This proves that a much higher degree of entanglement has been established in our final state.\\

\subsection{Teleportation}
We next tested whether the improved quality of ibm\_cairo over earlier processors allows MQNC to achieve its intended purpose of generating a state over which quantum information can be transferred. For each of the pairs in the quantum network code in turn, the graph state is generated using the LC procedure introduced earlier, and the MQNC measurements that generate the pair are performed. The state to be teleported is then encoded in a qubit adjacent to the first qubit of the pair, entangled with the first qubit of the pair, and the Pauli-X measurements to teleport it onto the second qubit are performed, as shown in Fig.~\ref{fig:mqncteleport} for the case of cross pair generation. 

Results are post-selected based on MQNC and teleportation measurement outcomes. Quantum process tomography is performed on this circuit and a Choi matrix representation of the teleportation channel is obtained \cite{Stormer2013}. We then use the Choi matrix to obtain the fidelity of sending different states through the teleportation channel. The standard basis is used, with 2000 shots per circuit. We collect 30 Choi matrices for each channel to obtain a mean and error bar on the extracted fidelity.

In the current context, the teleportation can be viewed as a game, where Alice wishes to communicate a quantum state to Bob without physically sending the state, and where Alice is allowed to perform a measurement on the state. A natural question which arises is whether it is possible to do so using only classical communication and no quantum correlations. While it is clearly impossible to gain enough information from a state with a single measurement to be able to perfectly recreate the state after classical transmission of the information, it is nonetheless possible to gain some information. This bound on the fidelity of classical `teleportation' is studied in Refs. \cite{Massar1995} and \cite{Massar2005}. For there to be any benefit to quantum teleportation and for confirming that quantum correlations have been used, the fidelity must exceed this classical bound \cite{Massar2005, Horodecki1996}. \\

The usual measure for teleportation fidelity is the average gate fidelity between the teleportation channel and the identity gate, defined according to the integral \(F_{ave}(\epsilon, I) = \int d\psi \bra{\psi}I^{\dagger}\epsilon(\ketbra{\psi})I\ket{\psi},\)
which runs over the entire Bloch sphere. Using this measure of the teleportation fidelity we find that only one cross pair (qubits 0 \& 5) and one straight pair (qubits 0 \& 1) have fidelity exceeding $2/3$, the average which can be achieved with classical communication alone without quantum correlations \cite{Horodecki1996}. 

We ask whether it is possible to guarantee better than classical results for all pairs if we have additional information about the identity of the states. This problem has been considered by Roy et al.~\cite{Roy2022} who showed that the increased information about the states to be teleported obtained by restricting them to a polar cap of the Bloch sphere  (see Fig.~\ref{fig:bloch} (a)) leads to an increase in the bound on the maximum fidelity obtainable by purely classical means. Here we look to exploit this and see if the gain in the average teleportation fidelity due to considering a small portion of the Bloch sphere having states that are transferred with better quality in our MQNC implementation is enough to raise it above the classical bound for all four pairs of the network code. The portion of the Bloch sphere we consider is a polar cap, as in Ref.~\cite{Roy2022}, but where the pole is rotated to coincide with the teleported state that has maximum fidelity for each pair and the fidelity of states near it are generally higher than states from the remainder of the Bloch sphere.\\

Instead of considering the reconstructed Choi matrix, $\Lambda$, for each channel using a pair, we use the Choi matrix to plot the teleportation fidelity for individual states on the surface of the Bloch Sphere in Fig.~\ref{fig:bloch} (b) for the four different teleportation routes.  Here, the fidelity is given by $F(\theta,\phi)=\bra{\psi}\epsilon(\rho_{in})\ket{\psi}$, with $\rho_{in}=\ketbra{\psi}$, $\epsilon(\rho_{in})={\rm Tr}_1[\Lambda(\rho_{in,1}^T\otimes I_2)]$ and $\ket{\psi}=\cos (\theta/2) \ket{0}+e^{i\phi}\sin (\theta/2) \ket{1}$. From the fidelity plots one can see that in all cases there are well defined areas of states which are teleported with relatively high fidelity, and that for each pair, these areas are stable over the entire 30 runs of the demonstration. Most of the high-fidelity states fall within a spherical cap (i.e. a rotated polar cap) of the Bloch sphere. 

\begin{figure}
  \includegraphics[width=0.5\textwidth]{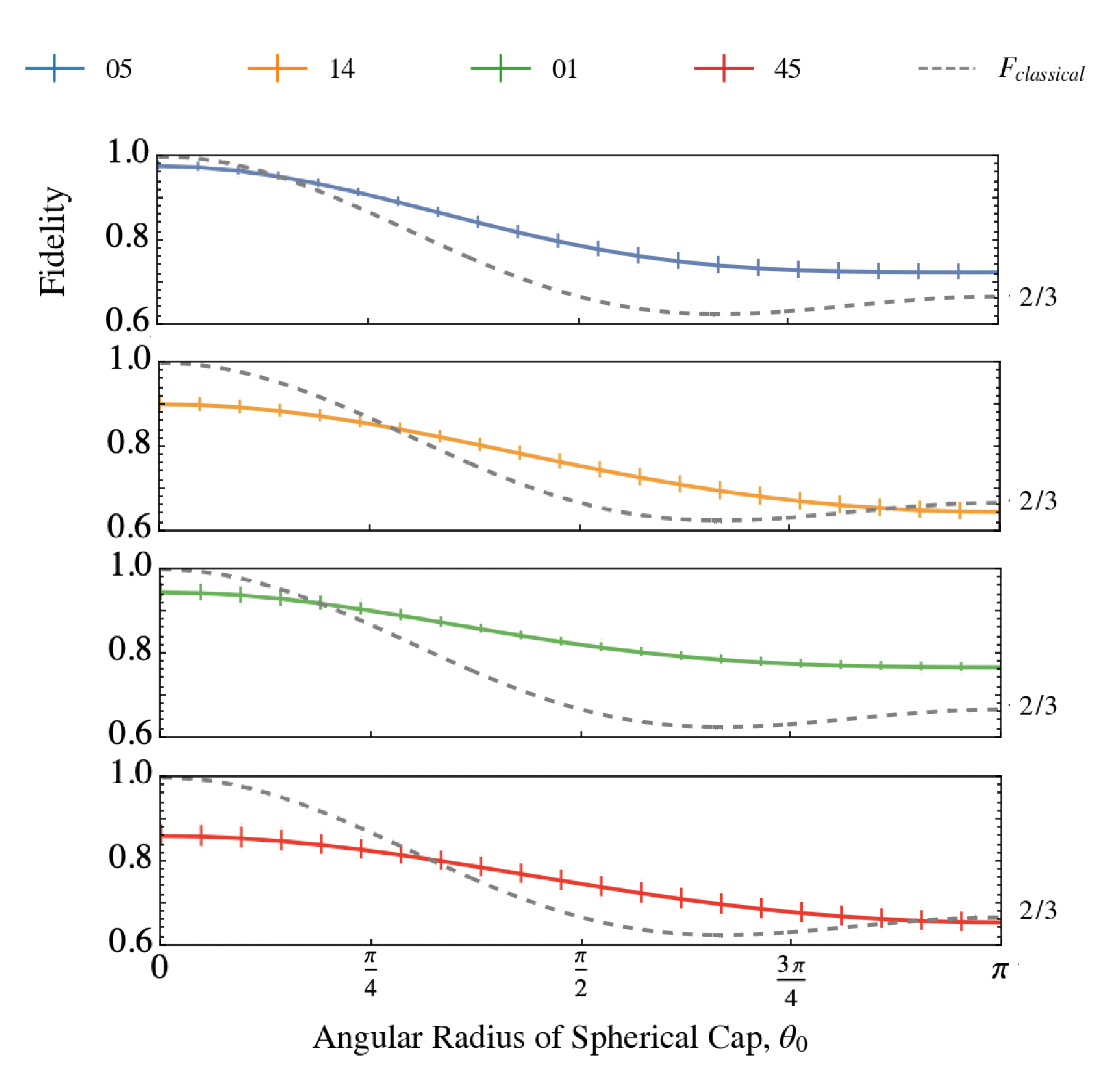}
  \caption{The average teleportation fidelity of states from a spherical cap of the Bloch sphere. Error bars correspond to one standard deviation of the fidelity obtained from 30 reconstructed Choi matrices for each channel. The case where $\theta_0 = \pi$ corresponds to the average over the entire sphere. Dashed lines show the highest fidelity achievable through classical communication alone.}
  \label{fig:spherical}
\end{figure}

For each pair of the network code we determine the point of highest teleportation fidelity, $F_{max}$, and calculate the average fidelity by performing numerical integration over pure states falling within a certain distance of this point. The distance is determined so as to give states within a spherical cap of angular radius $\theta_0$. This is repeated for a range of angles. For each of the 30 channels obtained (for each pair) the average fidelity at a given angular radius is calculated by numerical integration as \(F_{ave}(\epsilon, I,\theta_0) = \frac{1}{S}\int_0^{\theta_0} \int_0^{2 \pi} F(\theta,\phi) \sin \theta d \theta d \phi \), where $S= \int_0^{\theta_0} \int_0^{2 \pi} \sin \theta d \theta d \phi$. To take into account the rotated polar cap, the input state $\ket{\psi}$ is rotated by $\theta_{max}$ and $\phi_{max}$, before and after the channel, to align the state $\ket{0}$ along the axis of maximum fidelity, $F_{max}$, as shown in Fig.~\ref{fig:bloch} (b) for the four different teleportation routes.
One can see from Fig.~\ref{fig:spherical} that for all pairs the classical bound~\cite{Roy2022}, $F_{cl}(\theta_0)=1-[(2+\cos \theta_0)(1-\cos \theta_0)]/6$, is exceeded for a large range of $\theta_0$. This demonstrates that quantum correlations were used in the transfer of the quantum states over all four channels using MQNC.

It is interesting to note that the average teleportation fidelity in Fig.~\ref{fig:spherical} is different to the fidelity of the actual entangled pairs used, as shown in Fig.~\ref{fig:statetomo}. In the case of the latter, the fidelity is a fixed value quantifying the link quality, while the former can change depending on the size and shape of the distribution taken from the Bloch sphere. A higher value of the average teleportation fidelity than the entangled pair fidelity does not represent a contradiction because they represent two different aspects in terms of the channel quality. As the distribution is reduced around the point of maximum fidelity of the Bloch sphere (from Fig.~\ref{fig:bloch}), it is expected that the average fidelity of the smaller distribution will increase, and its value is less impacted by the fidelity of the entangled link. 

An interesting example is the extreme case, where the unentangled state $\ket{0}\ket{+}$ is used as the pair, which has a fidelity of 0.5 with respect to the 2-qubit graph state. However, when using it to teleport states near to the state $\ket{0}$, {\it i.e.} a small polar cap, the average teleportation fidelity approaches unity. Further details on the average teleportation fidelity for the polar cap distribution, and its relation to the entangled pair fidelity and concurrence can be found in the theory paper by Roy et al., in Ref.~\cite{Roy2022}.
\begin{figure*}[t]
\centerline{\includegraphics[width=\textwidth]{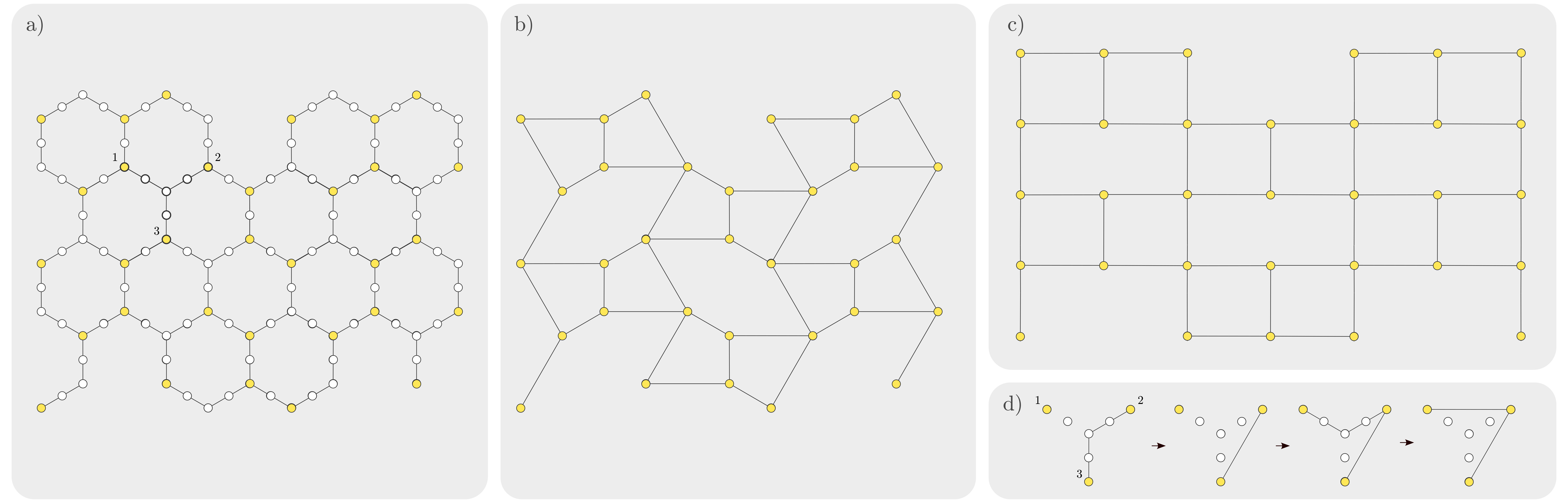}}
\caption[Embedding of the MQNC switch in the IBM processors.]{Embedding of the \(4\times 4\) MQNC switch in the IBM processors. (a) The heavy-hex layout. Switch qubits are represented as yellow dots. White dots represent auxiliary qubits used in the creation of the switch. Edges represent nearest-neighbour relations. (b) The MQNC switch graph state. Edges represent entanglement. (c) The same graph state as in (b), rearranged to show the grid-like layout. (d) The procedure for forming additional connections to a qubit.}
\label{fig:tessellation}
\end{figure*}

\subsection{Noise}
Our limited success in the preceeding subsection raises questions regarding the sources of error, and what may be needed to overcome these so that arbitrary states may be teleported. Unfortunately, the current state of the IBM Quantum processors is such that errors are highly variable to the point where even hourly recalibration does not result in consistent behaviour over a span of minutes. Since we are primarily concerned here with a repeatable demonstration that MQNC is already somewhat feasible on these processors, we have compensated for the variable errors by taking averages of results spanning multiple recalibrations. It makes sense then to consider an averaged error model when discussing such demonstrations. 

In Ref.~\cite{Rall2022}, we explored the efficacy of a depolarizing noise model for our MQNC implementation, and were able to predict the density matrices of the MQNC qubit pairs with $\sim 90\%$ fidelity. This suggests that a depolarizing channel model of noise represents well the collective impact of the error sources on the protocol in the IBM processor, complementing the results of Ref.~\cite{Pathumsoot2020} and serving as a starting point for ballpark estimates or comparison to similar demonstrations on IBM processors.

On the other hand, predictions regarding performance of future quantum processors are better made with reference to improvements of the physical hardware and consequent reduction in specific kinds of error. We can reasonably expect that single and two-qubit gate errors, readout errors, and T1 and T2 decoherence all contribute to the total error which is propagated to the final teleported state, so it would be sensible to study the expected effects of error in a controlled way using QISKIT's noisy QASM simulator. Given the small number of qubits involved, this is a reasonable proposition in terms of computational resources required, but since the analysis of such a study is non-trivial this is left for future work.

\section{Mapping the MQNC switch onto IBM Q processors}
We have demonstrated teleportation over MQNC on a single butterfly network, giving an indication of the degree of feasibility of MQNC on superconducting quantum processors at present. Considering the high degree of noise present even in this simple case, implementation of the MQNC switch is unfortunately not yet practical. Eventually, error correction will mitigate the effects of noise either completely or to a large extent, but in the interim we may still see significant reduction in noise which would enable implementation of the MQNC switch with some degree of success even in the absence of error correction. This would be of interest from the standpoint of demonstrating the feasibility of the MQNC switch and potential NISQ applications using it, and therefore we discuss how the switch may be embedded in a processor utilising the heavy-hex layout.\\

In Fig. \ref{fig:tessellation} we present an embedding of the MQNC switch in the heavy-hex layout (shown explicitly in Fig. \ref{fig:tessellation}(a), that makes use of a subset of the available qubits which forms a tessellating pattern shown in Fig. \ref{fig:tessellation}(b). This pattern is identical to the one in Fig. \ref{fig:tessellation}(c) and the generalised MQNC previously shown in Fig. \ref{fig:switch}(a). The qubits used are not adjacent, and so it is necessary to use graph state rewiring to entangle them. Given two qubits which are connected via an isolated path in the heavy-hex layout, the two qubits may be connected directly by entangling the qubits of the path via \(CZ\)'s and subsequently measuring all the qubits in the path apart from the ends in the \(Y\)-basis - following transformation rule T2. The majority of the connections shown in Fig. \ref{fig:tessellation} (b) are created in this way by performing \(Y\)-measurements on the white qubits in Fig. \ref{fig:tessellation} (a). It is however not possible to create all the required connections using a single round of \(Y\)-measurements, since not all white qubits lie on isolated paths. The qubits labelled 1,2, and 3 in Fig. \ref{fig:tessellation} (a), are an example of this. In order to connect qubit 2 to both qubit 1 and qubit 3, the connections must be formed in two separate \(Y\)-measurement steps, as shown in Fig. \ref{fig:tessellation}(d). Using this procedure it becomes possible to have up to four connections per qubit despite the heavy-hex layout permitting at most three direct connections per qubit. 

Note that the qubits which are measured out could potentially be used again to verify the presence of the desired connections using a parity checking method similar to that in \cite{Mooney2021}, with only minor adjustments being required to translate their GHZ-state technique to our graph states - at the cost of added complexity. This appears promising, but will likely require some optimization to be used effectively, and so we leave an investigation of it for future work where it may be treated with appropriate subtlety.

\section{Discussion}
Small-scale quantum communication networks will likely become feasible within the next few years as the technology continues to improve, however the NISQ regime presents challenges in the form of limited channel capacity and a limited number of qubits so that efficient use of these resources and their characterisation is paramount~\cite{Tham22,Buessen22}. 

The question of how practical MQNC might turn out to be in a realistic quantum communication network depends on many different aspects. For instance, when the physical resources are limited in a network, and multiple qubits are available for performing purification, these qubits could be used directly for solving contention. Furthermore, without a high-quality quantum memory available at network nodes and with probabilistic remote entanglement generation, the cost of pre-generation of the resource state could be very large. In addition, the topology of the network could limit the shape of the MQNC resource state in the multiparty case and one would need to adapt the MQNC multiparty state we have proposed in Section II C to that of the network topology, and determine whether the adapted state can be transformed into independent EPR pairs for teleportation between desired nodes, which is an NP-complete problem~\cite{Dahlberg20}. Thus, MQNC may provide an efficient and scalable solution to contention-free switching in quantum networks, potentially increasing network throughput compared to routing, but this depends on a number of practical aspects, including the allowed network topology. 

In this paper we presented the first successful demonstration of MQNC combined with quantum teleportation exceeding the maximum fidelity achievable by classical means. Our work builds on a prior demonstration wherein MQNC was studied on an older IBM superconducting processor whilst introducing a number of innovations. 
By introducing a rewiring scheme based on local complementation we are able to create the MQNC resource state on newer IBM processors whereas standard SWAP-based transpiling on this topology introduces too much noise for creation of the state. Using the new processors we were able to establish sufficient quantum correlations for teleportation of quantum information. We found however that the average teleportation fidelity for states sampled from the entire Bloch sphere only exceeded the classical bound for teleportation across one of the two final routes generated using MQNC (in either the crosswise or straight configuration). Some states were teleported with disproportionally low fidelity and the average teleportation fidelity increased considerably when taken over a distribution excluding these states. 

By adapting recent theory work on teleportation from non-uniform distributions we were able to show that the classical bound was exceeded in the case of a spherical cap distribution even when the increased classical knowledge about the state is accounted for. This likely extends to any teleportation scheme with biased noise. We also introduced a generalization of MQNC to an arbitrary number of source-destination pairs which requires few connections per qubit and is therefore suitable for 2D quantum processor architectures.\\
\begin{table*}[t]
\begin{tabular}{lrrrrrrrr}
Qubit & 2 & 3 & 5 & 8 & 9 & 11 & 14 & 13\\
\hline
T1 (\(\mu\)s) & 9.03\(\times 10^{1}\) & 1.64\(\times 10^{2}\) & 1.02\(\times 10^{2}\) & 9.74\(\times 10^{1}\) & 8.30\(\times 10^{1}\) & 7.43\(\times 10^{1}\) & 8.16\(\times 10^{1}\) & 1.11\(\times 10^{2}\)\\
T2 (\(\mu\)s) & 7.26\(\times 10^{1}\) & 1.95\(\times 10^{2}\) & 6.63\(\times 10^{1}\) & 7.85\(\times 10^{1}\) & 5.21\(\times 10^{1}\) & 1.25\(\times 10^{2}\) & 1.11\(\times 10^{2}\) & 1.78\(\times 10^{2}\)\\
Frequency (Hz) & 4.91\(\times 10^{9}\) & 5.12\(\times 10^{9}\) & 5.05\(\times 10^{9}\) & 4.97\(\times 10^{9}\) & 5.23\(\times 10^{9}\) & 5.13\(\times 10^{9}\) & 5.04\(\times 10^{9}\) & 5.28\(\times 10^{9}\)\\
Anharmonicity (Hz) & -3.44\(\times 10^{8}\) & -3.40\(\times 10^{8}\) & -3.41\(\times 10^{8}\) & -3.43\(\times 10^{8}\) & -3.40\(\times 10^{8}\) & -3.42\(\times 10^{8}\) & -3.42\(\times 10^{8}\) & -3.39\(\times 10^{8}\)\\
Readout error & 1.16\(\times 10^{-2}\) & 7.90\(\times 10^{-3}\) & 7.40\(\times 10^{-3}\) & 1.25\(\times 10^{-2}\) & 3.58\(\times 10^{-2}\) & 7.30\(\times 10^{-3}\) & 6.90\(\times 10^{-3}\) & 6.80\(\times 10^{-3}\)\\
Prob meas. 0 prep. 1 & 1.40\(\times 10^{-2}\) & 1.00\(\times 10^{-2}\) & 8.80\(\times 10^{-3}\) & 1.74\(\times 10^{-2}\) & 3.18\(\times 10^{-2}\) & 9.20\(\times 10^{-3}\) & 7.80\(\times 10^{-3}\) & 7.60\(\times 10^{-3}\)\\
Prob meas. 1 prep. 0 & 9.20\(\times 10^{-3}\) & 5.80\(\times 10^{-3}\) & 6.00\(\times 10^{-3}\) & 7.60\(\times 10^{-3}\) & 3.98\(\times 10^{-2}\) & 5.40\(\times 10^{-3}\) & 6.00\(\times 10^{-3}\) & 6.00\(\times 10^{-3}\)\\
Readout length (s) & 7.32\(\times 10^{-7}\) & 7.32\(\times 10^{-7}\) & 7.32\(\times 10^{-7}\) & 7.32\(\times 10^{-7}\) & 7.32\(\times 10^{-7}\) & 7.32\(\times 10^{-7}\) & 7.32\(\times 10^{-7}\) & 7.32\(\times 10^{-7}\)\\
ID error & 1.606\(\times 10^{-4}\) & 1.937\(\times 10^{-4}\) & 2.354\(\times 10^{-4}\) & 2.938\(\times 10^{-4}\) & 2.525\(\times 10^{-4}\) & 2.038\(\times 10^{-4}\) & 1.644\(\times 10^{-4}\) & 2.010\(\times 10^{-4}\)\\
\(\sqrt{X}\) error & 1.606\(\times 10^{-4}\) & 1.937\(\times 10^{-4}\) & 2.354\(\times 10^{-4}\) & 2.938\(\times 10^{-4}\) & 2.525\(\times 10^{-4}\) & 2.038\(\times 10^{-4}\) & 1.644\(\times 10^{-4}\) & 2.010\(\times 10^{-4}\)\\
Pauli-\(X\) error & 1.606\(\times 10^{-4}\) & 1.937\(\times 10^{-4}\) & 2.354\(\times 10^{-4}\) & 2.938\(\times 10^{-4}\) & 2.525\(\times 10^{-4}\) & 2.038\(\times 10^{-4}\)4 & 1.644\(\times 10^{-4}\) & 2.010\(\times 10^{-4}\)\\
\hline
Qubit pair & (2 3) & (3 5) & (5 8) & (8 11) & (8 9) & (11 14) & (14 13) & \\
\hline
\(CX\) error & 8.89\(\times 10^{-3}\) & 4.51\(\times 10^{-3}\) & 5.55\(\times 10^{-3}\) & 8.54\(\times 10^{-3}\) & 1.37\(\times 10^{-2}\) & 9.24\(\times 10^{-3}\) & 4.48\(\times 10^{-3}\) & \\
\end{tabular}
\caption{Calibration parameters from daily calibration of ibm\_cairo as reported by IBM Quantum. Last update date - 27 April 2022 at 9:10 AM (UTC).}
\label{table:cairo}
\end{table*}

\begin{table*}[t]
\begin{tabular}{lrrrrrrl}
Qubit & 0 & 1 & 5 & 6 & 10 & 11 & \\[0pt]
\hline
T1 (\(\mu\)s) & 1.230\(\times 10^{2}\) & 6.954\(\times 10^{1}\) & 7.037\(\times 10^{1}\) & 7.713\(\times 10^{1}\) & 1.123\(\times 10^{2}\) & 7.378\(\times 10^{1}\) & \\[0pt]
T2 (\(\mu\)s) & 8.261\(\times 10^{1}\) & 1.066\(\times 10^{1}\) & 5.538\(\times 10^{1}\) & 6.279\(\times 10^{1}\) & 6.371\(\times 10^{1}\) & 6.221\(\times 10^{1}\) & \\[0pt]
Frequency (GHz) & 5.020\(\times 10^{0}\) & 4.903\(\times 10^{0}\) & 5.143\(\times 10^{0}\) & 5.033\(\times 10^{0}\) & 4.960\(\times 10^{0}\) & 5.230\(\times 10^{0}\) & \\[0pt]
Readout error & 6.000\(\times 10^{-2}\) & 4.400\(\times 10^{-2}\) & 4.300\(\times 10^{-2}\) & 3.100\(\times 10^{-2}\) & 1.020\(\times 10^{-1}\) & 3.600\(\times 10^{-2}\) & \\[0pt]
Prob. meas. 0 prep. 1 & 6.200\(\times 10^{-2}\) & 8.000\(\times 10^{-2}\) & 5.200\(\times 10^{-2}\) & 5.200\(\times 10^{-2}\) & 1.780\(\times 10^{-1}\) & 4.000\(\times 10^{-2}\) & \\[0pt]
Prob. meas. 1 prep. 0 & 5.800\(\times 10^{-2}\) & 8.000\(\times 10^{-3}\) & 3.400\(\times 10^{-2}\) & 1.000\(\times 10^{-2}\) & 2.600\(\times 10^{-2}\) & 3.200\(\times 10^{-2}\) & \\[0pt]
\hline
Qubit pair & (0 1) & (0 5) & (1 6) & (5 6) & (5 10) & (6 11) & (10 11)\\[0pt]
\hline
CX error & 3.52\(\times 10^{-2}\) & 4.99\(\times 10^{-2}\) & 3.43\(\times 10^{-2}\) & 2.29\(\times 10^{-2}\) & 2.64\(\times 10^{-2}\) & 1.65\(\times 10^{-2}\) & 3.39\(\times 10^{-2}\)\\[0pt]
\end{tabular}
\caption{Calibration parameters from ibm Q20 Tokyo on 29 August 2019 at 7:00 PM (UTC) as reported by IBM Quantum.}
\label{table:tokyo}
\end{table*}

Our results show that high-fidelity teleportation of quantum states using MQNC on a real superconducting processor is possible, but only if the states belong to a subset of the Bloch sphere. Nonetheless, we have shown that it is already possible to transfer quantum information over the MQNC, which opens up the possibility of using entanglement purification to obtain a high-fidelity communication channel if several noisy ones are available \cite{Bennett1996, Deutsch1996}. Furthermore, there exist protocols such as the BB84 QKD protocol \cite{Bennett2014} which rely on repeat-until-success strategies and can therefore make use of an imperfect communication channel. Lastly, we note that while current processors are still too noisy for truly practical MQNC, the incorporation of error correction is planned for future IBM processors \cite{Chamberland2020}. This has the potential to negate the majority of the effects of noise in the protocol and there is the added advantage of the logical topology supporting direct creation of the MQNC switch via encoded graph states. The fact that present-day superconducting quantum processors are already capable of implementing butterfly MQNC with a relatively high degree of success suggests that MQNC presents a practical solution to contention in future NISQ quantum communication networks. 

\subsection*{Acknowledgments}
We acknowledge the use of IBM Quantum services for this work. The views expressed are those of the authors, and do not reflect the official policy or position of IBM or the IBM Quantum team. We thank Taariq Surtee and Barry Dwolatzky at the University of Witwatersrand and Ismail Akhalwaya at IBM Research Africa for access to the IBM processors through the Q Network and Africa Research Universities Alliance. This research was supported by the Department of Science and Innovation (DSI) through the South African Quantum Technology Initiative (SA QuTI), Stellenbosch University, the National Research Foundation (NRF), and the Council for Scientific and Industrial Research (CSIR).

\section*{Appendix}
The demonstration was carried out on ibm\_cairo version 1.0.24 between 9PM 26 Apr 2022 and 5AM 27 Apr 2022 (UTC). Table I lists calibration parameters for this processor during the daily calibration performed by IBM Quantum on the 27th of April 2022 at approximately 8-9AM (UTC). The layout of the processor is given in Fig.~\ref{fig:layout}. For the sake of rough comparison we include Table II, which lists IBM Q20 Tokyo properties at around the time of the experiment of Pathumsoot et al., but this is of limited relevance since the need for rewiring means that the improvements in ibm\_cairo do not directly translate to improvements in MQNC performance.

\end{document}